\documentclass[11pt,a4paper]{JHEP3}
\pdfoutput=1
\usepackage{epsfig}
\usepackage{amsmath}

\title{Two-dimensional gauge dynamics and the topology of singular determinantal varieties}

\author{Kenny Wong\\Department of Applied Mathematics and Theoretical Physics, \\Centre for Mathematical Sciences, \\University of Cambridge, \\Cambridge, CB3 0WA, UK\\{\tt k.wong@damtp.cam.ac.uk}}

\abstract{We record an observation about the Witten indices in two families of gauged linear sigma models: the U(2) model for linear sections of Grassmannians, and the U(1) model for quadric complete intersections. We describe how the Witten indices are related to the Euler characteristics of the singular skew-symmetric or symmetric determinantal varieties featuring in the analysis of their opposite phases, and we discuss the extent to which these relationships can be reconciled with standard Born-Oppenheimer arguments.}

\begin{document}
\maketitle
\flushbottom

\section{Introduction}

\paragraph{}
In this short note, we comment on the Witten indices in two families of gauged linear sigma models with quadratic potentials: the $U(2)$ model for complete intersections of sections of the determinant line bundle on Grassmannians, introduced in \cite{horitong}, and the $U(1)$ model for complete intersections of quadrics in projective space, studied in depth in \cite{twisted}.

\paragraph{}
These two families of GLSMs are known to realise dualities between certain pairs of non-birationally-equivalent varieties. The $U(2)$ model, for example, provides a physical interpretation of a duality between a pair of Calabi-Yau three-folds: a complete intersection in $\mathbf{Gr}(2,7)$, and a Pfaffian variety in $\mathbf P^6$; the two Calabi-Yau three-folds are believed to share the same mirror dual \cite{rodland}. The $U(1)$ model realises, among other examples, a duality between a pair of K3 surfaces: a complete intersection of three quadrics in $\mathbf P^5$, and a double cover over $\mathbf P^2$ branched over the degeneracy locus of a $6 \times 6$ symmetric form. These dualities can be interpreted as equivalences of derived categories \cite{borisov,kuzquadric, willed}, and the derived equivalences are examples of ``Homological Projective Duality'' \cite{kuzhpd,kuzgrass}.

\paragraph{}
In both the examples mentioned, a determinantal variety is involved somewhere in the construction: for the $U(2)$ model, this is the Pfaffian locus itself, whereas for the $U(1)$ model, it is the branching locus for the double cover. Crucially, these determinantal varieties are smooth. When one tries to generalise the analysis of these models to higher-dimensional varieties, these determinantal loci become singular, and the dynamics becomes more complicated. It is no longer possible to give a geometrical interpretation of the GLSMs in simple terms; such GLSMs are instead thought to describe ``non-commutative resolutions'' \cite{twisted,dbraneed,vandenbergh}.

\paragraph{}
Another kind of difficulty emerges when one attempts to apply the physical analysis of \cite{horitong} for the $U(2)$ model to Calabi-Yau varieties of even, rather than odd, dimension. The analysis of \cite{horitong} relies on a Born-Oppenheimer approximation, which is valid in odd dimensions but breaks down in even dimensions.

\paragraph{}
The purpose of this note is to record a simple observation about the Witten indices in these kinds of GLSMs, for varieties of \emph{arbitrary} dimension. Even though the relevant determinantal loci may be singular, and even though Born-Oppenheimer arguments may not be reliable, we nonetheless find that the Witten indices of these theories are related to the Euler characteristics of these determinantal loci and their loci of singular points in a regular and prescribed fashion, and in a way that bears great resemblance to Born-Oppenheimer arguments. For Calabi-Yau models in the $U(2)$ family, these patterns follow from results in \cite{efunction}, and for all other cases, the arguments in \cite{efunction} generalise in a simple way.

\paragraph{}
In Sections 2 and 3, we describe these properties of the Witten indices in each of our two families of GLSMs, commenting on their relationship to Born-Oppenheimer arguments. In Appendix A, we briefly sketch derivations of the results along the lines of \cite{efunction}. In Appendix B, we outline a small calculation used to interpret the results in Sections 2 and 3: the counting of quantum Coulomb branch vacua in $U(2)$ theories. Some mathematical details and computations in specific models are collected in Appendix C. 

\section{The $U(2)$ model: Pfaffians and Grassmannians}

\paragraph{}
The first of the models that we shall discuss is the two-dimensional $\mathcal N = (2,2)$ supersymmetric $U(2)$ gauge theory with $n$ chiral multiplets $\phi^\alpha_1, ... , \phi^\alpha_n$ in the fundamental representation and $k$ chiral multiplets $p^1, ... , p^k$ in the det$^{-1}$ representation of $U(2)$. The theory has a superpotential of the form,
\begin{eqnarray}
W = p^a A_a^{ij} \epsilon_{\alpha \beta} \phi^\alpha_i \phi^\beta_j,
\nonumber
\end{eqnarray} 
where each of $A_1^{ij}, \dots , A_k^{ij}$  is a skew-symmetric form in $\wedge^2 \mathbf C^n$. The theory has a Fayet-Iliopoulos parameter $r$ for the diagonal $U(1) \subset U(2)$.

\paragraph{}
For generic choices of $A_a^{ij}$, the $r > 0 $ phase of the GLSM is a smooth complete intersection of $k$ sections of the determinant line bundle $\wedge^2  (\mathcal S^\vee) $  in $\mathbf{Gr}(2,n)$. (See Appendix C for our conventions on bundles.) More explicitly, the $\phi^\alpha_i$ scalars acquire a vacuum expectation value proportional to $\sqrt{r}$, and parametrise the Grassmannian $\mathbf{Gr}(2,n)$. The superpotential sets $p^a = 0$ and cuts out the complete intersection,
\begin{eqnarray}
X = (A_a^{ij} \epsilon_{\alpha \beta} \phi^\alpha_i \phi^\beta_j = 0,   \ a = 1, ... , k) \subset \mathbf{Gr}(2,n).
\nonumber
\end{eqnarray}

\paragraph{}
The Witten index of the $r > 0$ phase is its Euler characteristic, $\chi(X)$.

\paragraph{}
The $r < 0$ phase is more complicated. The $p^a$ fields acquire a VEV proportional to $\sqrt{|r|}$, and become homogeneous coordinates for a $\mathbf P^{k-1}$. Identifying $[p^a] \in \mathbf P^{k-1}$ with the form $A^{ij}(p) := p^a A_a^{ij}$, the $\mathbf P^{k-1}$ can be viewed as the linear system spanned by our $k$ chosen sections of $\wedge^2 (\mathcal S^\vee)$ on $\mathbf{Gr}(2,n)$. For a fixed  $p^a$, the form $A^{ij}(p)$ acts as a complex mass matrix for the $\phi$ scalars. We split the analysis of this phase into two cases: $n$ odd and $n$ even.

\subsection*{Case: $n$ odd}
\paragraph{}
When $n$ is odd, $A^{ij}(p)$ is of rank $n - 1$ for generic $[p^a] \in \mathbf P^{k-1}$. Within this $\mathbf P^{k-1}$ is a nested sequence of closed subvarieties,
\begin{eqnarray}
\mathbf P^{k-1} = Y^1 \supset Y^3 \supset Y^5 \supset Y^7 \dots
\nonumber
\end{eqnarray}
where each $Y^d$ is the locus on which ${\rm rk}\left( A^{ij}(p) \right) \leq n - d$. Equivalently, $Y^d$ is the vanishing locus of the Pfaffians of the $(n - d + 1) \times (n - d + 1)$ minors of $A^{ij}(p)$. From a physical perspective, $Y^d$ is the locus of choices of values of $[p^a] \in \mathbf P^{k-1}$ setting at least $d$ linear combinations of the $\phi$ multiplets to be massless. Each $Y^d$ has codimension $d(d-1)/2$ in $\mathbf P^{k-1}$, so for $d(d-1)/2 > k-1$ (or for $d > n-1$), $Y^d$ is empty.

\paragraph{}
For generic choices of $A_a^{ij}$, all of the $Y^d$ loci are singular except for the final non-empty $Y^d$ in the sequence; furthermore, as will be explained in Appendix C, the locus of singular points of $Y^d$ is precisely the locus $Y^{d+2}$.

\paragraph{}
Suppose that one is to analyse the quantum dynamics of the $r < 0$ phase using a Born-Oppenheimer approximation. This is a two-stage process. First, one analyses the dynamics of the $\phi$ multiplets, treating the $p$ multiplets as fixed background fields. (One does this for all possible background values for $p$.) In this first stage, one is mainly concerned with the low-energy dynamics of $\phi$, and in particular, one wishes to count the number of supersymmetric vacua for $\phi$ and how this varies for different choices of $p$. In the second stage, one integrates out the $\phi$ multiplets, leaving an effective theory for the $p$ multiplets; the vacuum structure for the $\phi$ multiplets at different values of $p$ determines the local character of this new effective theory for $p$.

\paragraph{}
Let us apply Born-Oppenheimer to the $ r < 0$ phase of our GLSM. For a fixed background value of $p$, the local theory for the $\phi$ multiplets is an $SU(2)$ gauge theory with superpotential $W = A^{ij}(p) \epsilon_{\alpha \beta} \phi_i^\alpha \phi_j^\beta$. (The gauge group is $SU(2)$ rather than $U(2)$ because the $p$ multiplets tranform in the ${\rm det}^{-1}$ representation of $U(2)$, and fixing a background value for $p$ breaks the diagonal $U(1) \subset U(2)$.) The dynamics of this local theory depends crucially on how many of the $\phi$ multiplets are massive or massless for the given choice of $p$. To be precise, when $[p^a] \in Y^d \backslash Y^{d +2}$, there are $d$ massless linear combinations of $\phi$ multiplets. The massive $\phi$ multiplets can be integrated out, leaving an $SU(2)$ theory with $d$ massless flavours. It is shown in \cite{horitong} that $SU(2)$ with $d$ massless flavours has Witten index  $\left \lfloor{ \frac {d - 1} 2 }\right \rfloor$.

\paragraph{}
For $(n, k) = (5,5), (7,7)$ and $(9,9)$, this information leads to a clear proposal for the geometry of the $r < 0$ phase \cite{horitong}. What is special about these particular low-dimensional examples is that $Y^5, Y^7, Y^9, \dots $ all vanish, so the nested sequence of Pfaffians is simply $\mathbf P^{k-1} = Y^1 \supset Y^3 \supset \emptyset$. Moreover $Y^3$ is smooth. For $ [p^a] \in Y^1 \backslash Y^3$, the local theory for the $\phi$ fields has Witten index zero, whereas for $[p^a] \in Y^3 $, the local theory has Witten index one. Hence the Born-Oppenheimer approximation suggests that, at low energies, the theory localises to $Y^3$, that is, the theory is a sigma model with target space $Y^3$. Indeed, for $(n, k) = (5,5), (7,7)$ and $(9,9)$, the $Y^3$ loci are Calabi-Yau varieties of the same dimension as $X$ (the target space for the $r > 0$ phase), and furthermore, $\chi(X) = \chi(Y^3)$.

\paragraph{}
Our main objective is to comment on the extent to which one may generalise this physical analysis for arbitrary $n$ and $k$, where $Y^3$ is no longer smooth, and where the sub-loci $Y^5, Y^7, Y^9, \dots $ are no longer empty. As explained above, the Witten index of the local theory is zero for $[p^a] \in Y^1 \backslash Y^3$, one for $[p^a] \in Y^3 \backslash Y^5$, two for $[p^a] \in Y^5 \backslash Y^7$, and so on. Applying Born-Oppenheimer naively, one would be tempted to speculate that the $r < 0$ phase is a sigma model whose target space is a smooth resolution of $Y^3$ of a certain form: this resolution would be a single cover over $Y^3 \backslash Y^5$, but over $ Y^5 \backslash Y^7$ it would be a fibre bundle whose fibres have Euler characteristic two, and over $Y^7 \backslash Y^9$ the Euler characteristic of the fibres would jump to three, and so on.

\paragraph{}
Furthermore, if $n \neq k$, the target space $X$ for the $r > 0$ phase is not Calabi-Yau. If $ n > k$, for instance, the Fayet-Iliopoulos parameter $r$ flows under the renormalisation group from the $r > 0$ phase in the UV to the $r < 0$ phase plus $\frac 1 2 (n-k)(n-1) $  gapped vacua on the quantum Coulomb branch in the IR. If $n < k$, there are instead $\frac 1 2 (k-n)(n-1) $ vacua on the quantum Coulomb branch, but the RG flow is reversed. (Mathematical readers may think of the number of Coulomb branch vacua as the difference in the number of exceptional objects in the derived categories of the respective spaces. See Appendix B for details of how the quantum Coulomb vacua are counted.)

\paragraph{}
Returning to our discussion of Born-Oppenheimer, if the above interpretation of the $r < 0$ phase were correct, it would lead to a precise prediction about the relationship between the Euler characteristic of the complete intersection target space for the $r > 0$ phase, $\chi(X)$, and the Euler characteristics of the smooth quasi-projective subvarieties\footnote{Complex algebraic varieties obey an inclusion-exclusion principle: for a complex quasi-projective variety $X$ and a closed subvariety in $Y \subset X$, we have $\chi(X) = \chi(X \backslash Y) + \chi(Y)$. Also, if $F \to X \to B$ is a fibre bundle, whose fibre $F$, base $B$ and total space $X$ are all complex algebraic varieties, and which admits a trivialising open cover of Zariski-open sets, then the respective Euler characteristics obey the multiplicative property $\chi(X) = \chi(F) \chi(B)$ (see for instance \cite{multiplicative}).} $Y^{d} \backslash Y^{d+2}$, valid for odd $n$:
\begin{eqnarray}
\chi(X)& = & \frac 1 2 (n-k) (n-1) \nonumber \\ &&   \ \  + \  0 \times \chi( Y^1 \backslash Y^3) + 1 \times \chi (Y^3 \backslash Y^5) + 2\times  \chi(Y^5 \backslash Y^7) + 3 \times \chi(Y^7 \backslash Y^9) + \dots
\nonumber \\
\label{grodd}
\end{eqnarray}
This result does indeed hold for all odd $n$ and all $k$. For the Calabi-Yau cases, with $n = k$, it follows directly from arguments in \cite{efunction}. As we will explain in Appendix A, these arguments can be generalised to non-Calabi-Yau cases too. (In Appendix C, we  describe an efficient strategy for computing these Euler characteristics in specific cases and list some examples for small $n$.)

\paragraph{}
What is remarkable is that, although the prediction in equation (\ref{grodd}) is valid, the physical interpretation above cannot be entirely accurate. The Born-Oppenheimer approximation is valid, but the geometric description of the effective theory for the $p$ multiplets, as stated above, cannot be correct, because no global smooth resolution of $ Y^3$ of the kind described exists in general; the derived category of the $r < 0$ phase, that is, its category of B-branes, is instead believed to be a \emph{non-commutative resolution} of $Y^3$ \cite{twisted,kuzgrass}.

\paragraph{} And yet, the relationship between the Witten index and the Euler characteristics of the loci $Y^d \backslash Y^{d+2}$ in equation (\ref{grodd}), as predicted by Born-Oppenheimer, still holds. This is interesting from the perspective of the gauge dynamics: even in the absence of a genuine geometrical target space description for $r < 0$, our physical Born-Oppenheimer intuition still appears to capture an essential aspect of the low-energy dynamics of the theory.

\subsection*{Case: $n$ even}
Using the same notation as before, we obtain a nested sequence of Pfaffian subvarieties,
\begin{eqnarray}
\mathbf P^{k-1} = Y^0 \supset Y^2 \supset Y^4 \supset Y^6 \dots
\nonumber
\end{eqnarray}

\paragraph{}
If one is to conjecture a formula analogous to (\ref{grodd}) for even $n$, one might initially consider a sum of terms of the form $\left \lfloor{ \frac {d - 1} 2 }\right \rfloor \times \chi(Y^d \backslash Y^{d+2})$, since $ \left \lfloor{ \frac {d - 1} 2 }\right \rfloor$ is the expression for the Witten index of $SU(2)$ with $d$ flavours as computed in \cite{horitong}. However, this expression for the Witten index is obtained by a certain limiting procedure, which is problematic when $n$ is even. In \cite{horitong}, the theory is deformed by giving twisted masses to the $\phi$ fields. (This shifts the supersymmetric vacua to the quantum Coulomb branch, where they can be counted more easily.) Having counted the vacua, the twisted masses are then sent to zero. But there is a subtlety with taking this limit when $n$ is even: as the twisted masses tend to zero, the quantum Coulomb branch develops a flat potential, giving rise to a further continuous family of vacua, parametrised by the scalar in the $SU(2)$ vector multiplet \cite{horitong}. The presence of the flat direction invalidates the Born-Oppenheimer approximation.

\paragraph{}
It is possible to define the Witten index of $SU(2)$ with $d$ massless flavours by choosing a different limiting procedure -- a procedure that incorporates, and regularises, the contribution from the non-compact $SU(2)$ Coulomb branch. In \cite{localisation}, the elliptic genus of  $SU(2)$ with $d$ massless flavours is computed using localisation. This quantity is defined as
\begin{eqnarray}
Z_{\rm EG} = {\rm Tr} (-1)^F q^{L_0} \bar q^{\bar L_0} y^{J_0} \prod_{i = 1}^{d} x_i^{K_i},
\nonumber
\end{eqnarray}
for the infra-red fixed point of the theory. Here $L_0$ and $\bar L_0$ are the Virasoro generators, $J_0$ is the left-moving $U(1)$ R-symmetry and $K_1, ... , K_d$ are Cartan charges for the $SU(d)$ flavour symmetry. Thus this elliptic genus is a character evaluated by deforming the theory by introducing holonomies for the left-moving R-symmetry and the flavour symmetry. In the limit  $y \to 1$, $q \to 0$, $x_i \to 1$, the elliptic genus reduces to the Witten index. Let us examine this limit. First, sending $ q \to 0 $, $x_i \to 1$, one finds \cite{localisation} (see also \cite{longflow} for a discussion of the pure $SU(2)$ case) that
\begin{eqnarray}
\lim_{x_i \to 1} \lim_{q \to 0} Z_{\rm EG} = \frac{(y^{\frac 1 2} + y^{\frac 3 2} + y^{\frac 5 2} + \dots + y^{d- \frac 1 2}) - y^{\frac 1 2}}{1 + y}.
\nonumber
\end{eqnarray}
The $1+y$ has the appearance of a geometric series summing over contributions from bosonic zero modes over the non-compact Coulomb branch. In the limit $y \to 1$, we approach the boundary of the region where this geometric series converges, yet its summed form remains well-defined in the $y \to 1$ limit and gives a half-integer result,
\begin{eqnarray}
\lim_{y \to 1} \lim_{x_i \to 1} \lim_{q \to 0} Z_{\rm EG} = \frac {d-1} 2.
\nonumber
\end{eqnarray}
Although the Born-Oppenheimer approximation is not valid, one may nevertheless  conjecture that these regularised Witten indices appear as coefficients in a relationship between the Euler characteristics of the complete intersection and of the determinantal strata, analogous to (\ref{grodd}).

\paragraph{}
Indeed, for $n$ even, the correct relationship between the Euler characteristics is
\begin{eqnarray}
\chi(X)& = & \frac 1 2 (n-k) (n-1) \nonumber \\ &&   \ \  - \  \frac 1 2 \times \chi( Y^0 \backslash Y^2) + \frac 1 2 \times \chi (Y^2 \backslash Y^4) + \frac 3 2 \times  \chi(Y^4 \backslash Y^6) + \frac 5 2 \times \chi(Y^6 \backslash Y^8) + \dots
\nonumber
\\
\label{greven}
\end{eqnarray}

\paragraph{}
The coefficients in (\ref{greven}) agree with the regularised expression for the local Witten index computed from the elliptic genus. The constant term $\frac 1 2 (n-k) (n-1)$ is zero in Calabi-Yau examples, as one would expect, and equals the number of quantum Coulomb vacua when $k = 0$ (see Appendix B). In all other non-Calabi-Yau examples, with $n \notin \{ 0, k \}$, there is a slight discrepancy between this constant term and the true signed count of quantum Coulomb vacua, $ (n - k) \times \left \lfloor{ \frac {n - 1} 2 }\right \rfloor $, since $\frac{n-1}2$ is half-integer when $n$ is even.

% \paragraph{}
% It is worth pointing out that, since $\chi(Y^0) = \chi (\mathbf P^{k-1}) = k $, (\ref{greven}) can be rewritten as
% \begin{eqnarray}
% \chi(X)& = & \frac 1 2 (n-k -1) n \nonumber \\ &&   \ \  + 0 \times \chi( Y^0 \backslash Y^2) + 1 \times \chi (Y^2 \backslash Y^4) + 2 \times  \chi(Y^4 \backslash Y^6) + 3 \times \chi(Y^6 \backslash Y^8) + \dots
% \nonumber
% \end{eqnarray}

\paragraph{}
Finally, let us comment on the $(n,k) = (6,6)$ case. Here, $X$ is a K3 surface of degree 14 in $\mathbf{Gr} (2,6)$ and $Y^2$ is a Pfaffian cubic fourfold; the dualities between these two varieties are explored in \cite{beauville,hassett,kuzcubic,thomas} among many other references. Since $\chi(Y^0) = \chi (\mathbf P^5) = 6$, the relationship (\ref{greven}) reduces to $\chi(X) = -3 + \chi(Y^2)$. This can also be understood physically from the point of view of the \emph{abelian} gauged linear model for the cubic fourfold, which flows to three Coulomb vacua plus a K3 conformal field theory in the infra-red (see for instance \cite{natalie}).

\section{The $U(1)$ model: Quadric complete intersections}

\paragraph{}
Let us now turn to our other model: $\mathcal N = (2,2)$ supersymmetric $U(1)$ gauge theory with $n$ chiral multiplets $\phi_1, \dots \phi_n$ of charge $+1$, $k$ chiral multiplets $p^1, \dots, p^k $ of charge $-2$, and a superpotential,
\begin{eqnarray}
W = p^a A_a^{ij} \phi_i \phi_j.
\nonumber
\end{eqnarray}
This time, $A_1^{ij}, \dots A_k^{ij}$ are \emph{symmetric} forms in ${\rm Sym}^2\mathbf C^n$.

\paragraph{}
The $r > 0$ phase is a non-linear sigma model on the complete intersection of $k$ quadrics in $\mathbf P^{n-1}$,
\begin{eqnarray}
X = (A_a^{ij} \phi_i \phi_j = 0,  \ a = 1, \dots , k ) \subset \mathbf P^{n-1}.
\nonumber
\end{eqnarray}

\paragraph{}
In the $r < 0$ phase, the $p$ fields again parametrise a $\mathbf P^{k-1}$, and we have a filtration of closed determinantal subvarieties of the form,
\begin{eqnarray}
\mathbf P^{k-1} = Y^0 \supset Y^1 \supset Y^2 \supset Y^3 \supset Y^4  \supset \dots
\nonumber
\end{eqnarray}
$Y^d$ is the codimension $d(d+1)/2$ locus on which  ${\rm rk} \left( A^{ij}(p) \right) \leq n - d$, or equivalently, the locus where at least $d$ of the $\phi$ fields become massless. As before, all but the last of the non-empty $Y^d$ loci are singular, and for generic choices of $A^{ij}$, the locus of singular points in $Y^d$ is equal to $Y^{d+1}$.

\paragraph{}
Once again, we discuss how one might attempt to apply Born-Oppenheimer to the $r < 0$ phase of the theory. For a fixed background value for $p$, the local theory for the $\phi$ fields is a $\mathbf Z_2$ orbifold with superpotential $W = A(p)^{ij} \phi^i \phi^j$. The $\mathbf Z_2$ acts by sending $\phi^i \mapsto - \phi^i$; it is the subgroup of $U(1)$ left unbroken when the $p$ fields acquire a VEV. If $[p^a] \in Y^0 \backslash Y^1$, then all $n$ of the $\phi$ multiplets are massless and can be integrated out. There are two gapped vacua: one in the untwisted sector and one in the twisted sector (though, as we will mention later on, the untwisted sector vacuum survives the $\mathbf Z_2$ orbifold only if $n$ is even). However, if $[p^a] \in Y^1$, then at least one linear combination of $\phi$ fields is massless and the Witten index is not well-defined.

\paragraph{}
This Born-Oppenheimer approximation is known to determine the $ r < 0$ geometry in a number of examples \cite{twisted} (see also \cite{mark,ericgerbes}). For illustration, we review the $(n,k) = (6,3)$ case. Here, the $r > 0 $ phase is the K3 surface defined as a complete intersection of three quadrics in $\mathbf P^5$. As for the more difficult $r < 0$ phase, only $Y^0$ and $Y^1$ are non-empty, so the filtration is $\mathbf P^2 = Y^0 \supset Y^1 \supset \emptyset$. $Y^1$ is a sextic curve in $\mathbf P^2$. Since the local theory for $\phi$ has a pair of gapped vacua whenever $[p^a] \in Y^0 \backslash Y^1$, one expects that the appropriate geometry is a double covering over $Y^0 \backslash Y^1$. By examining the Berry phases of the pairs of gapped vacua around $Y^1$, the authors of \cite{twisted} show that $Y^1$ is a branching locus. Thus the $r < 0$ phase is a double cover of $Y^0$ branched over $Y^1$, and from this description it is clear that the Euler characteristics must obey the relation $\chi(X) = 2 \chi(Y^0 \backslash Y^1 ) + \chi ( Y^1)$.

\paragraph{}
Our main observation for this $U(1)$ model is a more general relationship between the Witten index and the Euler characteristics of the determinantal loci:
\begin{eqnarray}
\chi (X) & =  & (n-2k) + 1 \times \chi(Y^0 \backslash Y^1) + 2 \times \chi(Y^1 \backslash Y^2) + 1 \times \chi(Y^2 \backslash Y^3) + 2 \times \chi(Y^3 \backslash Y^4)  \nonumber \\ &&   \ \ \ \ \ \ \ \ \ \ \ \ \ \ \ \  + 1 \times \chi(Y^4 \backslash Y^5) + 2 \times \chi(Y^5 \backslash Y^6) + \dots \ \ \ \ ( n {\rm \ odd }), \nonumber \\ \nonumber
\\
\chi (X) & = & (n-2k) + 2 \times \chi(Y^0 \backslash Y^1) + 1 \times \chi(Y^1 \backslash Y^2) + 2 \times \chi(Y^2 \backslash Y^3) + 1 \times \chi(Y^3 \backslash Y^4)  \nonumber \\ &&    \ \ \ \ \ \ \ \ \ \ \ \ \ \  \ \ + 2 \times \chi(Y^4 \backslash Y^5) + 1 \times \chi(Y^5 \backslash Y^6) + \dots \ \ \ \ ( n {\rm \ even }). \nonumber \\ \label{quad}
\end{eqnarray}
This follows from a similar argument as for the previous model, and we will sketch this in Appendix A.

\paragraph{}
The $n - 2k$ term is simply the signed count of Coulomb branch vacua. The coefficients consisting of alternating ones and twos is more intriguing. As explained above, the local theory for the $\phi$ multiplets for a background value of $[p^a] $ in $ Y^d \backslash Y^{d-1}$ is a $\mathbf Z_2$ orbifold of a theory of $n$ free chiral multiplets, of which $n - d$ have a complex mass and $d$ are massless. For $d \geq 1$, the presence of these massless chirals invalidates the Born-Oppenheimer approximation. Yet the alternating ones and twos can be recognised as the Witten indices of a certain deformation of this $\mathbf Z_2$ orbifold: they are the Witten indices of a $\mathbf Z_2$ orbifold of $n - d$ chirals with complex mass and $d$ chirals  \emph{with twisted mass}:
\begin{eqnarray}
{\rm Tr } (-1)^F = \begin{cases} 1, \ \ \ \  n - d {\rm \ odd,} \\ 2, \ \  \ \  n - d {\rm \ even.} \end{cases} \nonumber
\end{eqnarray}
This formula is explained in \cite{horiduality}. The difference between the odd and even cases can be traced to an ambiguity in defining how the $\mathbf Z_2$  symmetry acts on the untwisted sector vacuum. The upshot is that the twisted sector vacuum always survives the $\mathbf Z_2$ quotient but the untwisted sector vacuum only survives the $\mathbf Z_2$ quotient if the number of flavours with complex mass is even.

\paragraph{}
The noncommutative resolution of $\mathbf P^{k-1}$ that describes the $r < 0$ phase of this abelian GLSM is the derived category of coherent $(\mathcal B_0 |_{\mathbf P^{k-1}})$-modules on $\mathbf P^{k-1}$, where $\mathcal B_0$ is the sheaf of even parts of  Clifford algebras on $\mathbf P({\rm Sym}^2 \mathbf C^n)$ \cite{kuzquadric}. The pattern of alternating ones and twos in (\ref{quad}) could be related to the fact that the standard representation of the even parts of the Clifford algebra ${\rm Cl}(\mathbf C^{n-d})$  decomposes into subrepresentations of odd and even degree when $n - d$ is even, but is irreducible when $n - d$ is odd. It would be interesting to explore this idea further.

\paragraph{}
In light of these numerical observations, it would be interesting to explore the possibility of extending the criteria of validity for the Born-Oppenheimer approximation in supersymmetric gauge theories, or in quantum field theories in general. We have seen two examples where the Born-Oppenheimer approximation appears invalid due to the presence of massless modes in the local theories, and yet, regularised versions of the Witten indices for these local theories nonetheless appear to capture the low-energy dynamics of the full theory, at least at the level of numerical relationships between Euler characteristics as expressed by formulas (\ref{greven}) and (\ref{quad}). This leads us to speculate that such behaviour is a more general feature of quantum field theories.

 \section*{Acknowledgements}
 The author would like to thank David Tong, Ed Segal, Natalie Paquette and Carl Turner for vital conversations, and Richard Eager and Anatoly Libgober for correspondences. The author is supported by Gonville and Caius College and the ERC Grant agreement STG 279943.

\appendix

\section{Appendix: Derivations}

\subsection*{The $U(2)$ model}

\paragraph{}
We now briefly sketch a derivation of (\ref{grodd}) and (\ref{greven}), following the reasoning in \cite{efunction}. Let us consider the incidence correspondence,
\begin{eqnarray}
Z = \left\{ ([\phi_i^\alpha],[ p^a]) \in \mathbf{Gr}(2,n) \times \mathbf P^{k-1}  \ \bigr\vert \  p^a A_a^{ij} \epsilon_{\alpha\beta} \phi_i^\alpha \phi_j^\beta = 0 \right\}.
\nonumber
\end{eqnarray}
Let $\pi_1 : Z \to \mathbf{Gr}(2,n)$ and $\pi_2 : Z \to \mathbf{P}^{k-1}$ be the natural projections. We evaluate $\chi(Z)$ in two ways, by considering $Z$ both as a fibration over substrata of $\mathbf{Gr}(2,n)$ and as a fibration over substrata of $\mathbf P^{k-1}$. In \cite{efunction}, the calculation is carried out for the Calabi-Yau case, $n = k$, but we will see that the arguments generalise for arbitrary $n $ and $k$.

\paragraph{}
First, consider the $\pi_1^{-1}$ fibre above a point $[\phi_i^\alpha] \in \mathbf{Gr}(2,n)$. If $[\phi_i^\alpha] \in X$, then this fibre is the whole of the $\mathbf P^{k-1}$, whereas if $[\phi_i^\alpha] \notin X$, then this fibre is a hyperplane $\mathbf P^{k-2}\subset \mathbf P^{k-1}$. This gives the expression,
\begin{eqnarray}
\chi(Z) & = & k \chi(X)+ (k-1) \left( \chi(\mathbf{Gr}(2,n)) - \chi(X)\right).
\nonumber
\end{eqnarray}
Now, consider the $\pi_2^{-1}$ fibre above a point $[p^a] \in Y^d \backslash Y^{d+2}$. This is the vanishing locus in $\mathbf{Gr}(2,n)$ of a section of $\wedge^2 (\mathcal S^\vee)$ of rank $n - d$. In \cite{efunction}, the Euler characteristic of this hypersurface is shown to be $\frac {1} 2 (d-1) + \frac 1 2 (n-1)^2$, giving a second expression,
\begin{eqnarray}
\chi(Z) = \sum_{d} \frac {d-1} 2 \chi(Y^d \backslash Y^{d+2}) + \frac {(n-1)^2} 2 \chi(\mathbf P^{k-1}),
\nonumber
\end{eqnarray} 
where the sum is over $d \equiv n {\rm \ mod \ } 2$ only. Combining the two expressions for $\chi(Z)$, and using the fact that $\chi(\mathbf{Gr}(2,n)) = \frac 1 2 n (n-1)$ (see Appendix C), we obtain our desired result,
\begin{eqnarray}
\chi(X) = \frac {(n-k)(n-1)} 2 + \sum_{d} \frac {d-1}2 \chi(Y^d \backslash Y^{d+2}).
\nonumber
\end{eqnarray}

\subsection*{The $U(1)$ model}

\paragraph{}
For the $U(1)$ model, the appropriate incidence correspondence to consider is
\begin{eqnarray}
Z = \left\{ ([\phi_i],[ p^a]) \in \mathbf P^{n-1} \times \mathbf P^{k-1}  \ \bigr\vert \  p^a A_a^{ij} \phi_i \phi_j= 0 \right\}.
\nonumber
\end{eqnarray}
By considering fibres above $\mathbf P^{n-1}$, we find, as in the previous case, that
\begin{eqnarray}
\chi(Z) = k \chi(X) + (k-1) (\chi(\mathbf P^{n-1}) - \chi(X)) .
\nonumber
\end{eqnarray}
Meanwhile, the fibre above a point $[p^a] \in Y^d \backslash Y^{d+1}$ is a quadric in $\mathbf P^{n-1}$ of rank $n-d$, which has Euler characteristic $\frac 1 2 (3 + (-1)^{n-d}) + (n-2)$ (see \cite{libgoberquadric}), giving
\begin{eqnarray}
\chi(Z) = \sum_d \frac 1 2 (3 + (-1)^{n-d}) \chi(Y^d \backslash Y^{d+1}) + (n-2)\chi(\mathbf P^{k-1}).
\nonumber
\end{eqnarray}
Combining the two expressions then gives our final result, equation (\ref{quad}),
\begin{eqnarray}
\chi(X) = (n-2k) + \sum_d \frac {(3 + (-1)^{n-d})} 2 \chi(Y^d \backslash Y^{d+1} ).
\nonumber
\end{eqnarray}

\section{Appendix: The Coulomb branch of $U(2)$ gauge theory}

\paragraph{}
Here, we compute the number of Coulomb branch vacua in the $U(2)$ gauge theory. The techniques are standard and straightfoward (see for instance \cite{horitong,morrison}), but we have not been able to find this result in the literature.

\paragraph{}
We first integrate out the chiral multiplets and the W-bosons, leaving an effective action for the vector multiplet $\sigma$. Up to gauge equivalence, $\sigma$ takes the diagonal form $\sigma = {\rm diag} (\sigma_1, \sigma_2)$. Integrating out $n$ fundamental chirals and $k$ det$^{-1}$ chirals induces an effective twisted superpotential for $\sigma$ of the form \cite{horitong,effaction,phases},
\begin{eqnarray}
\tilde W (\sigma) = - t(\mu) (\sigma_1 + \sigma_2 ) - n \sigma_1 \left( \log \frac {\sigma_1} \mu - 1 \right) - n \sigma_2 \left( \log \frac {\sigma_2} \mu - 1 \right) \nonumber \\ + k (\sigma_1 + \sigma_2) \log \left( \frac{- \sigma_1 - \sigma_2 }\mu  - 1 \right).
\nonumber
\end{eqnarray}
The RG flow of the complexified Fayet-Iliopoulos parameter $t(\mu)$ is given by \cite{phases},
\begin{eqnarray}
t(\mu) = (n - k) \log \frac \mu \Lambda,
\nonumber
\end{eqnarray}
where $\mu$ is the RG scale and $\Lambda $ is the cutoff scale.

\paragraph{}
Setting $\partial \tilde W / \partial \sigma_1 = \partial \tilde W / \partial \sigma_2 = 0$, we find that the Coulomb branch vacua are located at the solutions to the equations,
\begin{eqnarray}
\frac {\sigma_1^n}{(-\sigma_1 - \sigma_2)^k} = \frac {\sigma_2^n}{(-\sigma_1 - \sigma_2)^k} = \Lambda^{n-k}.
\nonumber
\end{eqnarray}
These equations have solutions whenever $\sigma_2 = \exp(2\pi q i / n) \sigma_1 $ for $q \in \mathbf Z_n$, and for each choice of $q$, the resulting equation for $\sigma_1$ is
\begin{eqnarray}
\sigma_1^{n-k } = ( - 1 - \exp (2\pi q i / n ) )^k \Lambda^{n-k},
\nonumber
\end{eqnarray}
which in turn has $|n-k|$ distinct solutions.

\paragraph{}
However, when counting the solutions, one should bear in mind that the effective twisted superpotential for $\sigma$ is only valid when the W-bosons, the $\phi$ fields and the $p$ fields are massive, that is, when
\begin{eqnarray}
\sigma_1 \neq \sigma_2, \ \ \  \ \  \sigma_1 \neq 0, \ \ \ \  \   \sigma_2 \neq 0, \ \ \ \  \  \sigma_1 + \sigma_2 \neq 0.
\nonumber
\end{eqnarray}
Hence the $q = 0$ solutions are invalid, and so are the $q= \frac n 2$ solutions arising when $n$ is even. Furthermore, the permutation $\sigma_1 \leftrightarrow \sigma_2$ is a Weyl transformation in the $U(2)$ gauge group, so it is only necessary to count solutions for $q$  in the range $0 < q< \frac n 2 $.

\paragraph{}
The conclusion is that the number of Coulomb branch vacua is
\begin{eqnarray}
|n - k| \times \left \lfloor{ \frac {n - 1} 2 }\right \rfloor  .
\nonumber
\end{eqnarray} 

\paragraph{}
(Note that the case when $n $ is even and $k$ is zero is slightly different: the restriction $\sigma_1 + \sigma_2 \neq 0$ does not apply, as there are no $p$ multiplets, and we find $n(n-1)/2$ quantum Coulomb vacua, agreeing with the Euler characteristic of $\mathbf{Gr} (2, n)$.)

\section{Appendix: Bundles, cohomology and tables}

\paragraph{}
In this appendix, we define our conventions for bundles on Grassmannians, summarise results about the cohomology of Grassmannians, expand on our characterisation of the singular loci of determinantal varieties, and provide some explicit computations of Euler characteristics relevant to our models.

\subsection*{Bundles on Grassmannians}

\paragraph{}
We first review some general properties of bundles on Grassmannians. The Grassmannian $\mathbf{Gr}(r,n)$ has a natural rank $r$ vector bundle $\mathcal S$, known as the tautological bundle: this is the sub-bundle of $\mathcal O^{\oplus n}$ whose fibre above a point $[V] \in \mathbf{Gr}(r, n)$ is precisely the $r$-plane $V \subset \mathbf C^n$ that the point represents. The tangent bundle of the Grassmannian is $\mathcal T_{\mathbf{Gr}(r,n)} = \mathcal S^\vee \otimes \mathcal Q$, where $\mathcal Q = \mathcal O^{\oplus n} / \mathcal S$ is the quotient bundle. In particular, the dimension of $\mathbf{Gr}(r,n)$ is $r(n-r)$. (See \cite{sharpegr1,sharpegr2,sharpegr3} for realisations of bundles on Grassmannians in GLSMs.)

\paragraph{}
The cohomology ring of $\mathbf{Gr}(r,n)$ is generated by Schubert cycles. These are in one-to-one correspondence with Young diagrams with at most $r$ rows and at most $n - r$ columns; a Young diagram with $m$ boxes corresponds to an element of $H^{2m}(\mathbf{Gr}(r,n))$. It follows that $\chi(\mathbf{Gr}(r,n)) = n!/r!(n-r)!$. The cup product in cohomology agrees with the usual Littlewood-Richardson tensor product for Young diagrams. Denoting the Schubert cycle corresponding to the Young diagram with $m_1, \dots,  m_k$ boxes in the first $k$ rows by $\sigma_{(m_1, \dots, m_k)}$, the Chern classes of the tautological and quotient bundles are
\begin{eqnarray}
c(\mathcal S) & = & 1 - \sigma_{(1)} + \sigma_{(1,1)} - \sigma_{(1,1,1)} + \dots + (-1)^r \sigma_{(1,1,...,1)},
\nonumber
\\
c(\mathcal Q) & = & 1 + \sigma_{(1)} + \sigma_{(2)} + \sigma_{(3)} + \dots + \sigma_{(n-r)}.
\nonumber
\end{eqnarray}

\subsection*{Complete intersections in Grassmannians}

\paragraph{}
The $r > 0$ phase for the $U(2)$ model has target space $X$, defined as the complete intersection in $\mathbf{Gr}(2,n)$ of a  section of the bundle $\mathcal E = \left(\wedge^2 (\mathcal S^\vee) \right)^{\oplus k}$. Provided that the section is chosen generically, this complete intersection is smooth. This follows by applying Bertini's theorem $k$ times to the line bundle $\wedge^2 (\mathcal S^\vee) $. (A version of Bertini's theorem states that the vanishing locus of a generic section of a basepoint-free line bundle on a smooth complex algebraic variety is smooth.)

\paragraph{}
We can compute the Euler characteristic of this complete interesection in particular cases by applying the Gauss-Bonnet theorem,
\begin{eqnarray}
\chi(X) = \int_X c ( \mathcal T_X) = \int_{\mathbf{Gr}(2,n)} c ( \mathcal E) \wedge c (\mathcal T_X).
\nonumber
\end{eqnarray}
Since $\mathcal E|_X$ is the normal bundle of $X$, the adjunction formula gives
\begin{eqnarray}
c( \mathcal T_X ) = \frac {c (\mathcal T_{\mathbf{Gr} (2,n)})}{c (\mathcal E)}. \nonumber
\end{eqnarray}
This is sufficient to express $ c( \mathcal T_X ) $ in terms of Schubert cells, for specific examples.

\subsection*{Pfaffian varieties}

\paragraph{}
We now turn to the Pfaffian varieties that feature in the analysis of the $r < 0$ phase. These form a filtration,
\begin{eqnarray}
\mathbf P^{k-1} = Y^1 \supset Y^3 \supset Y^5 \supset Y^7 \dots \ \ \ \ \ ( n {\rm \ odd}), \nonumber \\
\mathbf P^{k-1} = Y^0 \supset Y^2 \supset Y^4 \supset Y^6 \dots \ \ \ \ \ ( n {\rm \ even}),
\nonumber
\end{eqnarray}
where each $Y^d$ is the locus in $\mathbf P^{k-1}$ where the rank of the skew-symmetric form $p^a A_a^{ij}$ associated to the given point  $[p^a] \in \mathbf P^{k-1}$ is less than or equal to $ n - d$. As stated in the main text, each $Y^d$ has codimension $d(d-1)/2$ in $\mathbf P^{k-1}$, and if the $A_a^{ij}$ matrices are chosen generically, then the set of singular points in $Y^d$ is precisely $Y^{d+2}$. When $k = \frac 1 2 n (n-1)$, that is, when $\mathbf P^{k-1}$ is the \emph{complete} linear system of sections of $\wedge^2 (\mathcal S_{\mathbf{Gr}(2,n)}^\vee)$, this observation follows by simple linear algebra (see for instance \cite{morrison}). The $k < \frac 1 2 n (n-1)$ case then follows from the $k = \frac 1 2 n (n-1)$ case by successive applications of Bertini's theorem, viewing $\mathbf P^{k-1}$ as the intersection of $\frac 1 2 n (n-1) - k$ generic hyperplanes in $\mathbf P^{\frac 1 2 n(n-1) - 1}$.

\paragraph{}
The Euler characteristics of the Pfaffian strata $Y^d \backslash Y^{d+2} \subset \mathbf P^{k-1}$ can also be written in terms of Schubert cells, enabling us to compute them in particular cases. (The strategy we describe also works for symmetric determinantal varieties, with skew-symmetric tensor products replaced everywhere by symmetric ones. A similar strategy is used in \cite{morrison,morrison2} for a different class of determinantal varieties appearing in GLSMs; see also \cite{harris,pragacz} for general results about the topology of determinantal varieties.)

\paragraph{}
Since $Y^d$ is singular, we cannot apply Gauss-Bonnet directly, so we first compute the Euler characteristic of a resolution of $\tilde Y^d$, defined as the incidence correspondence,
\begin{eqnarray}
\tilde Y^d = \left((p,[V]) \in \mathbf P^{k-1} \times \mathbf{Gr}(n - d, n) \ \bigr\vert  \ A(p) \in \wedge^2 V \subset \wedge^2 \mathbf C^n \right) .
\nonumber
\end{eqnarray}
$\tilde Y^d$ is smooth, by a simple application of Bertini's theorem. Defining $\pi_1 : \tilde Y^d \to \mathbf P^{k-1}$ and $\pi_2 : \tilde Y^d \to \mathbf{Gr}(n - d, n)$ to be the natural projections, it is clear that
\begin{eqnarray}
(\pi_1)^{-1} (p) \cong \begin{cases} \emptyset  & {\rm if  \ }  p \notin Y^d; \\ \{ p \} & {\rm if \ } p \in Y^d \backslash Y^{d + 2}; \\ \mathbf{Gr}(d,d') & {\rm if  \ } p \in Y^{d'} \backslash Y^{d' + 2}  {\rm \  with \ } d' > d.\end{cases} \nonumber
\end{eqnarray}
% (More precisely, $\pi^{-1} (Y^{d'} \backslash Y^{d' + 2})$ is the Grassmannian bundle $\mathbf{Gr} ( d, \mathcal F_{d'})$ over $Y^{d'} \backslash Y^{d' + 2}$, where $\mathcal F_{d'}$ is the rank $d'$ subbundle of $(\mathcal O^{\oplus n})^\vee$ whose fibre above $p \in Y^{d'} \backslash Y^{d' + 2}$ is the annihilator of the unique $(n - d')$-plane $W \subset \mathbf C^n$ such that $A(p) \in \wedge^2 W$.)

% Furthermore, $\tilde Y^d$ is smooth. For $k = n(n-1)/2$, i.e. when $[A(p)^{ij}]$ ranges over all skew-symmetric forms in $\mathbf P (\wedge^2 \mathbf C^n)$ as $p$ varies over $\mathbf P^{k-1}$, the smoothness of $\tilde Y^d$ is clear, since $\tilde Y^d $ is isomorphic to the projective bundle $ \mathbf P(\wedge^2 \mathcal S)$ over $\mathbf{Gr} (n - d, d)$. For $k < n(n-1)/2$, $\tilde Y^d$ is then the intersection of the same $ \mathbf P(\wedge^2 \mathcal S)$  with $n(n-1)/2 - k$ sections of the basepoint-free line bundle $\pi_2^{\star} \mathcal O(1)$, and such an intersection is also smooth for generic choices of $A_a^{ij}$ by Bertini's theorem.

\paragraph{}
To evaluate the Euler characteristic of $\tilde Y^d$ by Gauss-Bonnet, we must describe $\tilde Y^d$ as the vanishing locus of a global section of a suitable vector bundle on $\mathbf P^{k-1} \times \mathbf{Gr}(n - d, n)$. Thinking of $A(p)$ as a global section of $\mathcal O_{\mathbf P^{k-1}}(1) \otimes \wedge^2 \mathbf C^{n} $ , the subvariety $\tilde Y^d \subset \mathbf P^{k-1} \times \mathbf{Gr} (n-d,n)$ is the locus where the pull-back $\pi_1^\star (A(p))$ vanishes as a section of the bundle,
\begin{eqnarray}
\mathcal E = \pi_1^\star \mathcal O_{\mathbf P^{k-1}}(1) \otimes \pi_2^\star \left( \frac {\wedge^2  \mathcal O^{\oplus n}_{\mathbf{Gr}(n-d,n)}}{\wedge^2 \mathcal S_{\mathbf{Gr}(n-d,n)}} \right).
\nonumber
\end{eqnarray}
Having expressed $\tilde Y^d$ as the vanishing locus of $\mathcal E$, the Chern class of $\mathcal E$ can now be expressed as a  polynomials in Schubert cells for $\mathbf{Gr}(n-d,n)$ and hyperplane sections of $\mathbf P^{k-1}$, and the Euler characteristic of $\tilde Y^d$ can be evaluated by applying the Gauss-Bonnet theorem as before.

\paragraph{}
Our real objective is to find the Euler characteristics of the differences, $Y^d \backslash Y^{d + 1} \subset \mathbf P^{k-1} $. We observed earlier that each $\pi_1^{-1} ( Y^{d'} \backslash Y^{d'+ 2} )$ is a $\mathbf{Gr}(d, d')$ bundle over $Y^{d'} \backslash Y^{d'+ 2} $. The $\mathbf{Gr}(d, d')$  fibre has Euler characteristic $d'!/d!(d'-d)!$, so we find that
\begin{eqnarray*}
\chi (\tilde Y^d) = \chi  ( Y^d \backslash Y^{d + 2} )  + \left(\frac{(d+2)!}{d!\times 2!} \right) \chi ( Y^{d+2} \backslash Y^{d + 4} )+ \left( \frac{(d+4)!}{d!\times 4!} \right)   \chi  ( Y^{d+4} \backslash Y^{d + 6} )+ \dots 
\nonumber
\end{eqnarray*}
(A similar decomposition technique is employed for other non-abelian GLSMs in \cite{johanna1, johanna2}.) 

\paragraph{}
Once we have obtained $\chi(\tilde Y^d)$ for all $d$, this relationship is sufficient for us to obtain $\chi(Y^d \backslash Y^{d+2})$ for all $d$. We list some low-dimensional examples in the tables below.

\newpage

\subsection*{Table of Euler characteristics: the $U(2)$ model}

% \paragraph{}
% As a preliminary set of examples, let us consider the case $k = n(n-1)/2$, where $\mathbf P^{k-1}$ is the \emph{complete} linear system of sections of $\wedge^2 (\mathcal S_{\mathbf{Gr}(2,n)}^\vee)$. The resolutions $\tilde Y^d$ are then simply the projective bundles $ \mathbf P(\wedge^2 \mathcal S_{\mathbf{Gr} (n-d, n) })$. Hence their Euler characteristics are
% \begin{eqnarray}
% \chi ( \tilde Y^d ) = \frac 1 2  (n-d)(n-d-1) \times \frac {n!}{(n-d)!\times  d!}.
%\nonumber
% \end{eqnarray}
% From this, it is easy to verify that
% \begin{eqnarray}
% \chi(Y^d \backslash Y^{d + 2} ) = \begin{cases} \frac {n (n-1)} 2, \ \ & d = n - 2 ;\\ 0 & {\rm otherwise,} \end{cases}\nonumber
% \end{eqnarray}
% for $k = n(n-1)/2$. These values clearly obey (\ref{grodd}) or (\ref{greven}) for all $n$, and provide a test of (\ref{grodd}) or (\ref{greven}) for cases where $Y^d$ is non-empty for $d$ arbitrarily high.

\begin{center}
\begin{tabular}{c | c | c c c c c}
$(n,k)$&  $\chi(X)$ & $\chi(Y^0 \backslash Y^2)$ &  $\chi(Y^1 \backslash Y^3)$ & $\chi(Y^2 \backslash Y^4)$ &  $\chi(Y^3 \backslash Y^5)$ &  $\chi(Y^4 \backslash Y^6)$  \\
\hline
(2,0) & 1 \\
(2,1) & & 1 \\
\hline
(3,0) & 3 \\
(3,1) & 2 && 1 \\
(3,2) & 1 && 2 \\
(3,3) & && 3 \\
\hline
(4,0) & 6 \\
(4,1) & 4 & 1 \\
(4,2) & 4 & 0 && 2 \\
(4,3) & 2 & 1 && 2 \\
(4,4) & 2 & 0 && 4 \\
(4,5) && 1 && 4 \\
(4,6) && 0 && 6 \\
\hline
(5,0) & 10 \\
(5,1) & 8 && 1 \\
(5,2) & 6 && 2 \\
(5,3) & 4 && 3 \\
(5,4) & 7 && $-1$ && 5 \\
(5,5) & 0 && 5 && 0 \\
(5,6) & 5 && $-1$ && 7 \\
(5,7) & && 3 && 4 \\
(5,8)&  && 2 && 6 \\
(5,9) & && 1 && 8 \\
(5,10) &&& 0 && 10 \\
\hline
(6,0) & 15 \\
(6,1) & 12 & 1 \\
(6,2) & 12 & $-1$ && 3 \\
(6,3) & 6 & 3 && 0 \\
(6,4) & 12 & $-5$ && 9 \\
(6,5) & $-6$ & 11 && $-6$ \\
(6,6) & 24 & $-21$ && 27 \\
(6,7) & $-14$ & 29 && $-36$ && 14 \\
(6,8) & 14& $-29$ && 51 && $-14 $ \\
(6,9) && 21 && $-36$ && 24 \\
(6,10) && $-11$ && 27 && $-6$ \\
(6,11) && 5 && $-6$ && 12 \\
(6,12) && $-3$ && 9 && 6 \\
(6,13) && 1 && 0 && 12 \\
(6,14) && $-1$ && 3 && 12 \\
(6,15) && 0 && 0 && 15 \\
\end{tabular}
\end{center}

\subsection*{Table of Euler characteristics: the $U(1)$ model}

% \paragraph{}
% Similar to the previous model, the cases  with $k = n(n+1)/2$ can be solved by simple means. Here, $\tilde Y^d$, the smooth resolution defined analogously to the previous case with skew-symmetric products replaced by symmetric products, is $\mathbf P({\rm Sym}^2 \mathcal S_{\mathbf{Gr}(n-d, n)})$, and its Euler characteristic is
% \begin{eqnarray}
% \chi (\tilde Y^d) = \frac 1 2 {(n-d)(n-d+1)}\times \frac{n!}{(n-d)! \times d!}.
% \nonumber
% \end{eqnarray}
% From this, we obtain, for $k = n(n+1)/2$,
% \begin{eqnarray}
% \chi(Y^d \backslash Y^{d+1}) = \begin{cases} n, \ \ & d = n - 1 ;\\ \frac {n(n-1)}{2}, & d = n-2; \\ 0 & {\rm otherwise.} \end{cases}
% \nonumber
% \end{eqnarray}
% This is consistent with equation (\ref{quad}).

\begin{center}
\begin{tabular}{c | c | c c c c c }
$(n,k) $ & $ \chi(X) $  & $ \chi(Y^0 \backslash Y^1 ) $  & $ \chi(Y^1 \backslash Y^2 ) $ & $\chi(Y^2 \backslash Y^3 ) $ & $\chi(Y^3 \backslash Y^4)$ & $ \chi( Y^4 \backslash Y^5) $ \\
\hline
(1,0) & 1 \\
(1,1) & 0 & 1 \\
\hline
(2,0) & 2 \\
(2,1) & 2 & 1 \\
(2,2) && 0 & 2 \\
(2,3) && 1 & 2 \\
\hline
(3,0) & 3 \\
(3,1) & 2 & 1 \\
(3,2) & 4 & $-1$ & 3 \\
(3,3) && 3 & 0 \\
(3,4) && $-1$ & 1 & 4 \\
(3,5) && 1 & 2 & 2 \\
(3,6) && 0 & 3 & 3 \\
\hline
(4,0) & 4 \\
(4,1) & 4 & 1 \\
(4,2) & 0 & $-2$ & 4 \\
(4,3) & 8 & 7 & $-4$ \\
(4,4) && $-10$ & 4 & 10 \\
(4,5) && 11 & 4 & $-10$ \\
(4,6) && $-10$ & 4 & 12 \\
(4,7) && 7 & $-4$ & $-4$& 8 \\
(4,8) && $-2$ & 4 & 6 & 0 \\
(4,9) && 1 & 0 & 4 & 4 \\
(4,10) && 0 & 0 & 6 & 4 \\
\hline
(5,0) & 5 \\
(5,1) & 4 & 1 \\
(5,2) & 8 & $-3$ & 5 \\
(5,3) & $-8$ & 13 & $-10$ \\
(5,4) & 16 & $-31$ & 15 & 20 \\
(5,5) && 55 & 0 & $-50$ \\
(5,6) && $-82$ & 1 & 87 \\
(5,7) && 106 & $-33$ & $-101$ & 35 \\
(5,8) && $-106$ & 73 & 111 & $-70$ \\
(5,9) && 82 & $-81$ & $-77$ & 85 \\
(5,10) && $-55$ & 55 & 60 & $-50$ \\
(5,11) && 31 & $-16$ & $-42$ & 22 & 16 \\
(5,12) && $-13$ & 3 & 26 & 4 & $-8$ \\
(5,13) && 3 & 2 & $-6$ & 6 & 8 \\
(5,14) && $-1$ & 1 & 2 & 8 & 4 \\
(5,15) && 0 & 0 & 0 & 10 & 5
\end{tabular}
\end{center}

\end{document}